# SEEING MORE, TREATING SMARTER: ROLE OF LONG-AXIAL FIELD-OF-VIEW PET/CT IN THE EVOLUTION OF THERANOSTICS

**Authors and Affiliations:**

**Pedro L. Esquinas, PhD**
Clinical Nuclear Medicine Physicist, Molecular Imaging and Therapy, BC Cancer, Adjunct Assistant Professor, Department of Radiology, , The University of British Columbia, Vancouver, Canada
pedro.esquinasfernandez@bccancer.bc.ca

**Fereshteh Yousefirizi, PhD**
Staff Scientist, Department of Integrative Oncology, BC Cancer Research Centre, Vancouver, Canada
Adjunct Assistant Professor, Department of Radiology, The University of British Columbia, Vancouver, Canada
Post-doctoral Research Fellow, Department of Basic and Translational Research, BC Cancer Research Institute, Vancouver, Canada
frizi@bccrc@bc.ca

**Ian Alberts, MBBS MA MD PhD FEBNM**
Nuclear Medicine Physician, Department of Molecular Imaging and Therapy, BC Cancer, Vancouver,
ian.alberts@ubc.ca

**Nicolas A. Karakatsanis, PhD**
Associate Professor of Biomedical Engineering, Department of Radiology, Weill Cornell Medicine, New York, New York, USA
nak2032@med.cornell.edu

**Arman Rahmim, PhD DABSNM**
Professor, Departments of Radiology, Physics and Biomedical Engineering, University of British Columbia, Canada
Distinguished Scientist and Provincial Medical Imaging Physicist at BC Cancer Research Centre, Vancouver, Canada
arman.rahmim@ubc.ca

**Carlos F. Uribe, PhD, MCCPM (Corresponding Author)**
Clinical Assistant Professor, Department of Radiology, The University of British Columbia, Vancouver, Canada
Leader of Clinical Nuclear Medicine Physics (Provincial Operations), BC Cancer, Vancouver, Canada
carlos.uribe@bccancer.bc.ca

## Key Points

- Long-axial field-of-view (LAFOV) PET/CT systems can deliver up to an order of magnitude (4x-40x) increase in sensitivity and enable streamlined whole-body dynamic imaging in a single bed position.
- Routine dynamic whole-body PET, which is more clinically adoptable with LAFOV systems, provides high-quality, time-resolved data for robust physiologically based pharmacokinetic (PBPK) modeling and predictive dosimetry, enabling truly personalized radiopharmaceutical therapies.
- Extended and multi-time-point PET imaging, coupled with LAFOV-enabled radiomics and AI methods, create the foundation for outcome prediction modeling and the development of digital twins.

## Synopsis

Long-axial field-of-view (LAFOV) PET/CT has the potential to redefine the role of molecular imaging in theranostics by making multiparametric whole-body (MPWB) imaging and predictive dosimetry more clinically feasible. Compared to conventional PET systems, LAFOV scanners provide dramatic gains in sensitivity and coverage, allowing dynamic acquisitions, delayed imaging, and dual-tracer protocols within clinically feasible workflows. These advances supply the quantitative data required for physiologically based pharmacokinetic (PBPK) modeling and the creation of theranostic digital twins, supporting true personalization of radiopharmaceutical therapy.

## Clinics Care Points

- LAFOV PET/CT has the potential streamlined MPWB imaging in a single bed position, improving adoption of dynamic and time-resolved protocols that provide deeper biological insights than conventional SUV-based scans.
- Enhanced sensitivity allows for delayed-time point and dual-tracer imaging, improving lesion detectability and enabling more confident therapy selection and monitoring.
- Quantitative data from LAFOV PET/CT supports PBPK modeling and predictive dosimetry, providing a path toward prospective, personalized treatment planning in radiopharmaceutical therapy.
- High-quality radiomics derived from LAFOV PET/CT scans offer reproducible biomarkers of tumor and organ heterogeneity, enriching prognostic modeling and guiding individualized patient management.
- By enabling digital twinning frameworks, LAFOV PET/CT fosters the integration of imaging, clinical, and biological data into computational models that simulate therapy response and optimize treatment strategies.

## Introduction

Positron Emission Tomography/Computed Tomography (PET/CT) plays a central role in staging and guiding the management of many malignancies [1], and its clinical utility continues to expand into a range of non-oncological indications. With the rise of theranostics [2], the role of PET raises an important question: could routine PET acquisitions be expanded to provide the data needed to predict accuracte patient-specific radiopharmaceutical therapy doses or even to construct digital twins that simulate therapy outcomes before treatment?

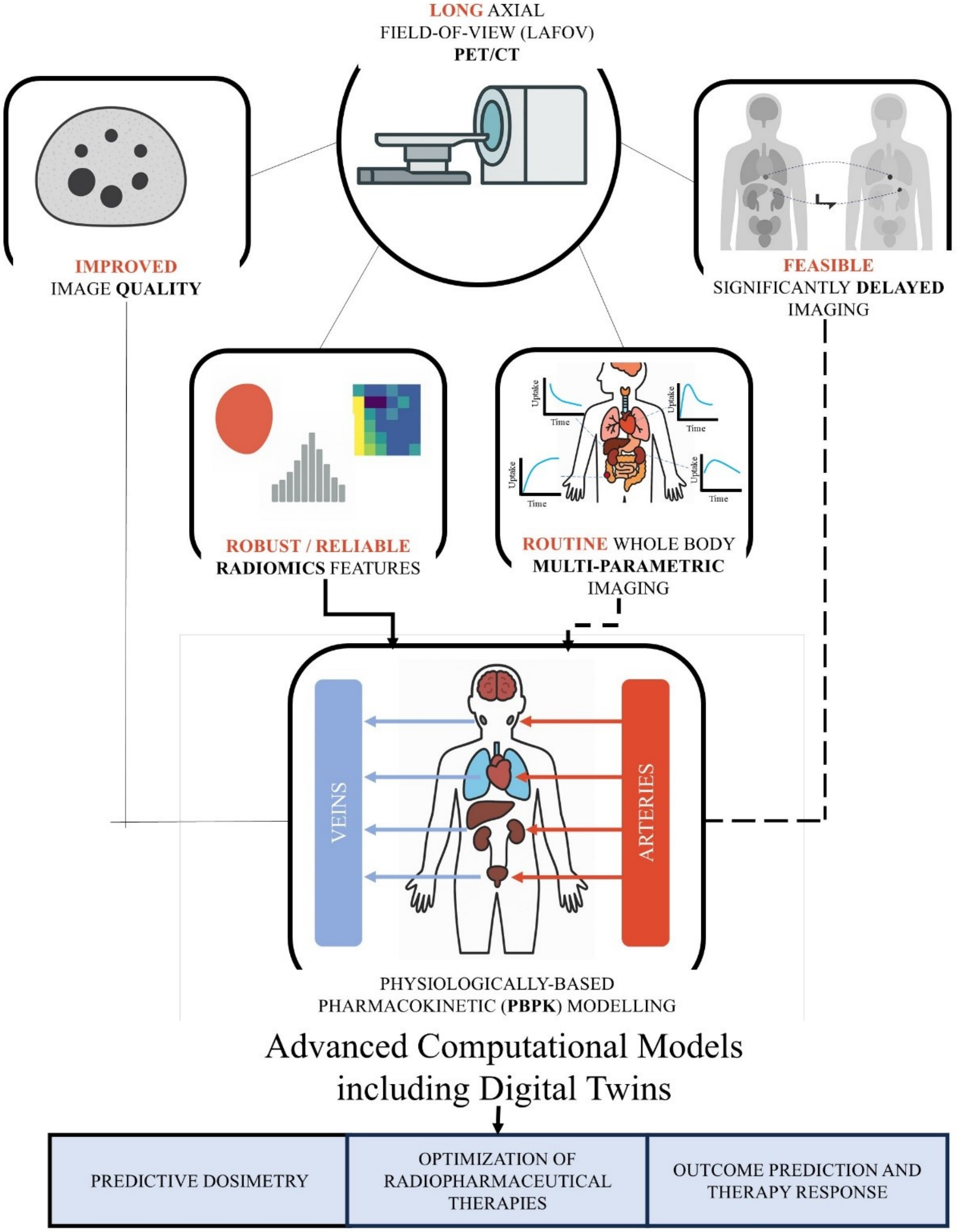

*Figure 1. An overview of interconnected advances related to LAFOV PET/CT, and how they collectively enable the next generation of personalized radiopharmaceutical therapies.*

Recent advances in scanner design have introduced a disruptive innovation to the field; long-axial field-of-view (LAFOV) PET/CT. Compared to conventional (short axial field-of-view, or SAFOV) PET systems with axial fields-of-view of ~20–30 cm, LAFOV scanners provide increased sensitivity due to their extended axial coverage and wider acceptance angles [3–6]. In addition to improved sensitivity, the extended axial coverage enables simultaneous whole-body (for most of the patients) static or dynamic imaging in a single bed position [6,7] as opposed to sequential limited AFOV scans at multiple overlapping bed positions over a single (static) or multiple (dynamic) passes [7–9], thus offering not only greater patient comfort but also workflow efficiency [10]. It is our view that the ability to simultaneously capture dynamic radiopharmaceutical distribution across the entire body represents a fundamental shift in PET imaging.

The performance gains from LAFOV PET are well documented [11]. What remains underexplored, however, is how these technical advances open new possibilities for theranostics [12]: delayed imaging once thought impractical, dynamic whole-body acquisitions that support physiologically based pharmacokinetic (PBPK) modeling, and multi-time-point or dual-tracer protocols that can refine patient selection and treatment optimization.

In this review, we explore how LAFOV PET/CT can transform theranostics (**Fig. 1**) by enabling multiparametric imaging, predictive dosimetry, and the eventual development of Theranostic Digital Twins.

## *Fundamentals of Multiparametric Whole-Body PET Imaging*

Multiparametric Whole-Body (MPWB) is an advanced dynamic imaging technique that extends traditional (static) PET by acquiring sequential scans over multiple time points, capturing the temporal behavior of radiopharmaceutical uptake post injection across organs and tissues and in the blood [11,12].

The key advantage of MPWB imaging lies in its ability to quantify tracer kinetics which describe how radiopharmaceuticals distribute, bind, and clear from tissue over post injection time. Kinetic modeling or linear graphical analysis, such as the Patlak method, can be applied to dynamic PET data to enable the generation of multiple parametric images each reflecting a specific biologically meaningful parameter of the tracer like, in the case of Patlak, the net uptake rate constant (Ki or slope) and the total blood distribution volume (intercept or DV) [13–15]. These parameters provide deeper physiological insights than static standardized uptake value (SUV) measurements and are less confounded by post injection time and the availability of the radiotracer in the blood which may change between exams [16–20].

**Table 1** compares SAFOV and LAFOV PET systems in the context of MPWB imaging. On SAFOV PET scanners, the MPWB protocol involves multiple rapid multi-bed passes of scanner across the body. Each pass contributes to a time–activity measurement for each voxel in tissue and in the blood regions eventually forming across all passes the tissue time-activity curves (TACs) for each tissue voxel and the blood plasma input function. Both are subsequently fitted to the Patlak model, enabling the robust voxel-wise estimation of the Ki and DV parameters forming the respective Ki and DV parametric images. The blood plasma input function can be estimated non-invasively by using image-derived or population-based methods eliminating the need for arterial sampling [21]. MPWB PET is also foundational for Physiologically Based Pharmacokinetic (PBPK) modeling, which mathematically describes radiopharmaceutical transport and interaction across organs [22] (in particular, see articles “An overview of PBPK and PopPK Models: Applications to Radiopharmaceutical Therapies for Analysis and Personalization” by Hardiansyah et.al and “Verification, Validation, and Uncertainty Quantification (VVUQ) of PBPK Models for Theranostic Digital Twins: Towards Reliable Model-Informed Treatment Planning for Radiopharmaceutical Therapies” by Zaid et.al of this issue on PBPK models and techniques to validate and deploy them [23,24]). Validated PBPK models may enable clinicians to simulate and predict therapeutic behavior in a patient specific way and MPWB PET is a big component in making this happen.

While MPWB PET has shown promise in oncology [16], inflammation [25], infection [26] , and evaluating systemic interactions such as the gut-brain [27] and heart-brain axes [28], widespread clinical adoption has been hindered by practical constraints. Among those are patient motion, long acquisition times (up to 90 minutes), and protocol and data complexities [9,27,28]. To overcome these barriers, LAFOV PET systems offer a transformative solution.

### *LAFOV PET as a Pathway to Routine MPWB Imaging and Personalized RPTs*

For MPWB PET to inform treatment planning, it must be scalable, streamlined and practical. SAFOV are capable of MPWB imaging but are limited by their short axial coverage at each bed position resulting in complex multi-pass multi-bed dynamic WB PET scan protocols with temporal gaps in the acquisition that are less practical and adoptable to the routine clinical workflow [9,29,30]. LAFOV PET scanners, on the other hand, cover the entire body (for a majority of the population) in a single bed position. This allows for simultaneous and continuous dynamic imaging of all relevant organs with greatly improved sensitivity and temporal resolution at a single pass which is a far more streamlined data acquisition protocol. These features result in high-quality parametric imaging by enabling: i) more reliable estimation of image-derived input functions (IDIF) from large blood pool regions that are always inside the extended AFOV, ii) more accurate tissue time-activity curves from less noisy and larger number of dynamic frames [31]. The combination of continuous and higher temporal resolution dynamic imaging of LAFOV PET acquisitions allows for higher sensitivity in the detection of kinetic characteristics therefore permitting the quantification of kinetic parameters within shorter scan time windows relative to SAFOV acquisitions. However, it should be noted that the kinetic parameters rely on biologic and pharmaco-kinetic effects that follow their own biological half-life which cannot be accelerated, hence there is a limit in how short the total scan time windows can be set. Recent studies [32] have demonstrated that even 10-min MPWB scans (e.g. at 55-65 minutes post-injection) may be sufficient for robust Patlak analysis of satisfactory accuracy, however significant errors were observed when further reducing the scan time. This matches existing clinical workflows for [$^{18}$F]FDG PET/CT and opens the door for routine implementation. Moreover, LAFOV PET provides data needed for the development and personalization of PBPK modeling, which is central to predictive dosimetry.

Despite this potential, several challenges remain for LAFOV routine clinical implementation. The extended axial coverage increases the likelihood of patient motion accumulating across the field-of-view; in contrast, SAFOV acquisitions are less impacted since motion artifacts are localized to the shorter coverage region. Similarly, the higher sensitivity and wide acceptance angles of LAFOV scanners increase the contribution of scatter and random events, particularly from very high-activity regions, which can degrade quantitative accuracy if not corrected adequately. Another practical concern is the proximity of multiple high-uptake regions within the same AFOV—for example, intense myocardial uptake adjacent to the aortic arch—where imperfect spillover correction may bias image-derived input function (IDIF) measurements. In addition to these physics- and physiology-related challenges, LAFOV PET generates vastly larger datasets than SAFOV systems, especially during dynamic MPWB acquisitions, requiring robust infrastructure with high-performance computing and large storage for efficient post-processing of data. Furthermore, MPWB PET requires dedicated, user-friendly software for the generation of parametric images, challenging its integration in routine clinical practice although vendor-specific solutions are increasingly available [33,34].

### *Predictive physiologically based pharmacokinetic (PBPK) Modeling and Prospective Dosimetry.*

True personalization of radiopharmaceutical therapies will require a shift from retrospective dosimetry towards prospective and predictive absorbed dose calculations (i.e., Treatment

Planning for radiopharmaceutical therapies). PBPK models support this personalization by simulating the biodistribution of radiopharmaceuticals based on physiological parameters.

Recent works from Kletting and Glatting are showing that physiologically based radio pharmacokinetic (PBRPK) modelling is required for predicting the administered activity in radiopharmaceutical therapies [35–40], which can accurately model the difference in biodistributions between the pre-therapeutic and therapeutic radiopharmaceutical distributions. Similarly, our own group's early data [41–43] already shows that PBPK modeling is feasible and highly informative, allowing us to study and analyze the biodistribution (i.e. time-activity curves) of therapeutic radiopharmaceuticals under different scenarios, facilitating experimentation that can lead to guidance for optimized prescribed activities and injection intervals. Our preliminary results also show that such modeling enables prediction of delivered doses from data even before the first RPT cycle [44].

In this context, MPWB PET plays a critical role in PBPK model parameter estimation by providing time-resolved quantitative data on radiopharmaceutical distribution across multiple organs. LAFOV PET further enhances this process compared to SAFOV PET by enabling continuous, simultaneous dynamic imaging of all relevant organs within a single field of view, providing richer, more complete datasets with higher temporal resolution and improved signal-to-noise ratio across the full kinetic profile. This reduces uncertainties in parameter estimation and enables more robust and individualized PBRPK models, leading to more accurate predictions of therapeutic radiopharmaceutical behavior and improved treatment planning for RPTs.

One recent study using LAFOV PET showed a comprehensive whole-body pharmacokinetic assessment of [$^{18}$F]PSMA-1007 [45], highlighting the modality's value for theranostics research. In another study, Ding *et al. 2023* [46] whole-body PBPK model of $^{68}$Ga[Ga]-NOTA-SGC8 was fitted to MPWB PET, predicting aptamer's pharmacokinetics in good agreement with clinically acquired data, offering a useful tool for more efficient disease diagnosis and lesion detection.

Even with LAFOV PET, key challenges for PBPK modeling remain. First, the high complexity and multi-compartmental nature of PBPK systems introduce numerous parameters, which can lead to poor identifiability and model overfitting when data are sparse or correlations exist between parameters. Second, inter-individual biological variability (e.g., organ volumes, blood flows, metabolic rates) requires personalized inputs, yet standardized physiological reference datasets are often lacking. Another challenge is patient motion, which can accumulate across the extended AFOV of LAFOV PET, propagating through the entire dynamic dataset and biasing both tissue TACs and input function estimates. Finally, the need for specialized modeling expertise presents an additional barrier to routine clinical adoption.

### *Beyond SUV: Predictive Modeling and Outcome Estimation.*

Having access to MPWB PET images enables the generation of parametric images that more directly quantify underlying biological parameters of interest, such as blood flow, glucose metabolism, and receptor binding [47]. Routine PET parametric imaging, which is streamlined and more clinically adoptable with LAFOV, introduces a new dimension of information about radiopharmaceutical behavior that goes beyond a single semi-quantitative value, such as SUV. The limitations of SUV [17–20] and the closely related metabolic tumor volume (MTV) are well established [48]. Previous studies have demonstrated that Patlak images derived from DWB 18F-FDG PET on SAFOV systems can improve malignant lesion sensitivity, accuracy, and potentially specificity compared to SUV-based PET imaging [49]. One would expect that such improvements in lesion detection seen with SAFOV MPWB PET would be further enhanced with LAFOV systems, owing to their increased temporal resolution and reduced noise in dynamic data. However, recent studies evaluating MPWB PET on LAFOV systems have reported that while Patlak Ki images provide better lesion contrast than SUV images, they do not significantly improve lesion detection rates in the clinic [50].

Although the advantages of MPWB PET for improved lesion detectability remain under investigation, we believe that dynamic and multi-parametric PET acquired prior to and during therapy will play an increasingly important role in dosimetry and outcome prediction, as it provides richer quantitative information than standard static SUV imaging that is less confounded to factors impacting accuracy across different exams hence resulting in more robust and reliable value quantitative comparisons between baseline and follow-up exams. Two main approaches can be considered to leverage this data. First, patient-specific PBPK models derived from routine LAFOV DWB PET (e.g., from MPWB $^{68}$Ga-PSMA scans) could enable accurate prediction of personalized therapy (e.g., $^{177}$Lu-PSMA) time-activity curves and allow simulation of different injection protocols, thereby facilitating true treatment planning optimization. Once the PBRPK model is determined for a given patient, therapy time-activity curves can be generated by simply substituting the diagnostic radionuclide with the therapeutic one. Second, with the capabilities of AI, it may become possible to directly predict organ and tumor dosimetry from pre-therapy MPWB PET data, bypassing the need for explicit PBPK parameter estimation. Several groups have already explored correlations between diagnostic PET metrics and post-treatment absorbed doses [51–53]. In addition, researchers have begun investigating the added predictive value of non-imaging biomarkers, such as estimated GFR [54], and others have demonstrated that combining pre-therapy PET imaging with clinicopathological biomarkers can predict $^{177}$Lu-RPT absorbed doses in tumors [55]. It is important to note that these studies have so far utilized static pre-therapy SUV PET images. We hypothesize that the richer information provided by MPWB PET, particularly when combined with non-imaging biomarkers, will further enhance the predictive performance of these models.

### *Optimized application of AI-based prospective dosimetry*

In addition to PBPK-based prospective dosimetry, artificial intelligence (AI) can be also applied for the prospective dosimetry based on pre-therapy PET [56,57], particularly in the prediction based on dynamic PET. LAFOV PET provides the unique opportunity for multi-organ or even whole-body dynamic PET imaging, enabling capture of dynamic uptake of most critical organs. This allows better application of AI methodology in bridging the missing pharmacokinetics in dosimetry prediction. Preliminary work based on computational simulation confirmed the potential of dynamic LAFOV PET in the prediction of post-treatment dosimetry.

### *Towards Theranostic Digital Twins*

The previous sections have argued that establishing routine MPWB PET in clinical practice will enable more accurate personalization of radiopharmaceutical therapies through the development of precise, patient specific PBPK models. In essence, this creates the foundation for building so-called Theranostic Digital Twins [58,59], with PBPK modeling playing a central role in this framework. A true digital twin should accurately represent radiopharmaceutical behavior, incorporating whole-body organ interactions and multi-organ kinetic models, and ultimately also include models of radiobiology to predict responses to therapies (see articles “Quantitative and Computational Radiobiology for Precision Radiopharmaceutical Therapies” by Yusufaly et.al and “Mathematical and Computational Nuclear Oncology: Toward Optimized Radiopharmaceutical Therapy via Digital Twins” by Ryhiner et.al of this issue on digital twins and radiobiological models [60,61]).

However, is the ability to perform routine MPWB PET on all patients sufficient to fully realize Theranostic Digital Twins? The answer is likely no. While dynamic and parametric PET imaging are frontier technologies, and while methods have been established [59], the problem of reliably estimating PBPK model parameters from MPWB PET data remains an active area of investigation. Beyond PBPK models, additional components are required for the construction of a comprehensive Theranostic Digital Twin, including the multi-parametric PET/CT images themselves, post-treatment SPECT/CT images, and relevant clinical biomarkers such as blood radiopharmaceutical concentrations, GFR, and other laboratory or clinical data.

This vision is compelling, but raises key questions: how can Theranostic Digital Twins be engineered, and what type of computational framework is best suited to integrate this diverse information? Such a model would need to incorporate imaging data, clinical variables, and mathematical kinetic models, effectively creating a highly multimodal system that captures both features and relationships between entities (e.g., tissues, organs, and temporal dynamics). Graph Neural Networks may represent a promising approach for modeling such complex structured data [62]. This area presents an exciting research direction for developing Theranostic Digital Twins; however, like many AI-based approaches, it will require large datasets. Given that MPWB PET is still limited in availability and that paired pre-treatment MPWB PET/CT and post-treatment SPECT/CT datasets are scarce, the use of simulated datasets will be crucial to initiate the development and study of Theranostic Digital Twins.

## *Delayed-time Point, Dual-tracer PET Imaging and Radiomics on LAFOV*

### *Delayed Time-point PET Imaging*

As discussed, LAFOV PET/CT systems are transforming the landscape of cancer imaging by enabling a new spectrum of capabilities [45,63,64]. Among these advances, dual time-point imaging has emerged as a particularly promising application. Dual time-point imaging (DTPI) over a long dynamic range offers a pragmatic alternative to full dynamic (MPWB) acquisitions for assessing radiopharmaceutical kinetics as simple metrics such as the retention index or dual-time-point Patlak analysis can be derived from two time points . These later time-points in the diagnostic scan enable better extrapolation of the radiopharmaceutical from diagnostic PET to therapeutic behavior of $^{177}$Lu. Furthermore, DTPI data can be leveraged for delta-radiomics, an emerging technique showing promise in predicting patient outcomes [65]. On the other hand, DTPI may be limited by the need to accurately co-register the early and late frames and the more difficult scheduling and implementation of exams involving two scan sessions per subject. In addition, DTPI is not robust to non-linear kinetic effects such as the expression of net efflux rate by the administered tracer which is an effect neglected by standard linear Patlak. In such scenario, the non-linear generalized Patlak analysis may be utilized instead, which is however not compatible with DTPI as it requires more than two temporal measurements [13,65].

We believe LAFOV PET will enable more sensitive patient selection for radiopharmaceutical therapies by making late-time point imaging feasible. Indeed, acquisition of images at later time points is associated with improved tumor-to-background ratio (TBR) and lesion detectability. For example, in [$^{18}$F]FDG imaging, scans performed at 2 hours post-injection have demonstrated superior performance in Deauville scoring relative to the standard 1-hour acquisition, prompting a re-evaluation of existing protocols for lymphoma [45]. Enhanced lesion visibility improves disease staging and restaging accuracy, enabling more confident clinical decision-making and personalized patient management. However, it is important to note that improved TBR at late time points might be tracer- and disease-specific. Future studies should therefore define which tracers and tumor types benefit most from late imaging

Despite these advantages, standard clinical adoption has been limited primarily due to the short half-life of radiopharmaceuticals such as $^{68}$Ga (68 min), resulting in substantial decay by later imaging time points. Additionally, prolonged acquisition times required to achieve adequate count statistics can limit scanner availability for other patients and are difficult to incorporate into busy clinical workflows [9]. Alberts et al. [66] demonstrated that using LAFOV PET systems for [$^{68}$Ga]Ga-PSMA-11 imaging, acquisitions performed up to four hours post-injection maintained high image quality, significantly improving TBR and signal-to-noise ratio (SNR), while requiring total acquisition times comparable to standard clinical protocols [5]. Thus, using LAFOV PET technology facilitates late imaging with sufficient image quality and statistics, and with increase TBR and SNR, leading to an increase in the sensitivity and effectiveness of patient selection for targeted radio-pharmaceutical therapies.

### *Dual-tracer protocols*

Multi-tracer imaging enabled by LAFOV PET/CT systems presents a transformative opportunity for theranostics that integrates molecular imaging for diagnosis, personalized treatment delivery, and monitoring of therapeutic response [12]. SAFOV PET/CT scanners are inherently limited by their restricted axial coverage, often requiring multiple bed positions to image the entire body. This not only increases scan time but may also compromise quantitative accuracy due to stitching artifacts and variable sensitivity across beds [63,67]. Moreover, implementing dual-tracer protocols with SFOV systems has traditionally been challenging due to concerns about radiation exposure and logistical demands, often necessitating multi-day imaging sessions [12,67]. The enhancement in sensitivity of LAFOV enables the execution of multi-tracer protocols with reduced scan durations and/or radiotracer doses [12,63].

For example, in mCRPC, combining [$^{68}$Ga]Ga-PSMA-11 and [$^{18}$F]FDG PET/CT is clinically valuable for assessing tumor heterogeneity and guiding eligibility for [$^{177}$Lu]Lu-PSMA therapy. LAFOV PET/CT facilitates this combination in a single imaging session, eliminating the need for staggered acquisitions across multiple days and addressing prior concerns about cumulative radiation burden [12,67]. This approach enables the detection of discordant lesions, such as those with high FDG uptake but low PSMA expression, which may inform prognosis and alter therapeutic strategy [12].

Similarly, dual-tracer protocols combining [$^{18}$F]FDG and Fibroblast Activation Protein Inhibitor (FAPI) agents have been explored to capture both metabolic activity and fibroblast activation within the tumor microenvironment. A feasibility study using total-body PET/CT demonstrated that a single session protocol—utilizing ultra-low-dose [$^{18}$F]FDG (0.37 MBq/kg) and half-dose [$^{68}$Ga]FAPI-04 (0.925 MBq/kg)—achieved high lesion detectability and led to clinically relevant changes in management for 44% of patients [12]. Notably, the overall radiation dose from this one-stop imaging protocol was comparable to that of a standard whole-body PET/CT scan using only full-dose [$^{18}$F]FDG, illustrating both the diagnostic value and patient safety advantages.

Altogether, the ability of LAFOV PET/CT systems to perform rapid, low-dose, and comprehensive multi-tracer imaging across the entire body marks a significant advancement in theranostics. This capability supports more accurate patient stratification, refined therapy selection, and precise response monitoring within a single imaging session [12].

## *Radiomics in LAFOV PET: Opportunities for Theranostics*

The term radiomics was first introduced in 2010 by Gillies et al. [68], defined as “the extraction of quantitative features from radiographic images.” These features capture subtle patterns and properties of tissues and lesions that may not be visually appreciable [69,70]. While early applications focused primarily on tumor characterization, recent studies have emphasized the prognostic and predictive value of features derived from non-tumor tissues and organs (e.g., fat, bone)[71,72] .

The integration of radiomics [70,73] with LAFOV PET/CT systems presents a new opportunity in theranostics for extracting robust imaging biomarkers and advancing personalized medicine. Radiomic features in PET/CT, which extend beyond conventional visual interpretation, provide deeper biological insights into tissue composition and lesion characteristics, supporting nuanced disease assessment. The high sensitivity and low intrinsic noise of LAFOV systems substantially improve the reliability and reproducibility of radiomics features. Most features achieve excellent agreement with full-count data at standard or extended acquisition times. Some features remain robust even under ultra-short or low-dose acquisition protocols, highlighting the feasibility of dose-reduction strategies without compromising quantitative integrity. Compared to standard-axial PET/CT, LAFOV systems demonstrate superior capability in characterizing both inter- and intra-patient heterogeneity. This is particularly valuable for therapy response assessment and risk stratification, where subtle differences in tissue or lesion heterogeneity may influence clinical decision-making. The reliability of radiomics features, even at very short acquisition times or with reduced radiotracer activity, makes LAFOV PET/CT particularly suitable for repeated or dynamic imaging. This also contributes to enhanced patient comfort and minimized radiation exposure.

Radiomics features extracted from LAFOV PET/CT scans using PyRadiomics [74] demonstrate high stability and reproducibility across multiple tracers and acquisition times, supporting their reliability for clinical and research use [75]. Overall, LAFOV radiomics enables the development of personalized, adaptive therapeutic strategies by capturing detailed disease characteristics and heterogeneity. These data-driven insights are essential for optimizing theranostic protocols and improving clinical outcomes.

## SUMMARY

Long-axial field-of-view PET/CT is transforming nuclear medicine by making routine multi-parametric whole-body imaging clinically adoptable and data-rich. The enhanced sensitivity and temporal resolution of LAFOV systems allow for precise PBPK modeling and prospective dosimetry, making personalization of radiopharmaceutical therapy more feasible and accurate. Moreover, the ability to perform delayed time-point imaging, alongside dual-tracer imaging and robust radiomics, sets the stage for digital twinning of patients and improved treatment planning. As these technologies mature and integrate into clinical workflows, they will fundamentally revolutionize how we select, plan, monitor and personalize radiopharmaceutical therapies.

## Disclosures

Arman Rahmim and Carlos Uribe are the co-founders of Ascinta Technologies. Pedro Esquinas is an employee at Ascinta Technologies. This research was funded by the Canadian Institutes of Health Research (CIHR) Project Grant (PJT-180251), Canadian Institutes of Health Research (CIHR) Project Grant (PJT-173231). All other co-authors have no relevant conflicts of interest.

## Tables

**Table 1.** Comparison of SAFOV and LAFOV PET/CT systems in the context of MPWB imaging.

| Workflow Aspect | SAFOV | LAFOV |
| --- | --- | --- |

| **Axial Coverage per bed position** | Limited coverage [20-30 cm]; requires multiple bed positions | Whole-body coverage (>= 1 m, depending on scanner); single bed position captures all organs of interest simultaneously |
|---|---|---|
| **Protocol for MPWB imaging** | Multiple rapid multi-bed passes; each pass contributes partial time-activity measurements | Single continuous dynamic acquisition of the entire body; streamlined protocol with no need for bed movement |
| **Temporal Sampling** | Temporal gaps between bed positions; lower temporal resolution | Continuous sampling across all organs simultaneouslyl ultra high temporal resolution achievable |
| **Tissue Time-Activity Curves** | Deroved from multiple passes; more susceptible to motion and temporal misalignment | Derive from continuous data; more robust, less noisy, less prone to motion misalignment |
| **Blood Plasma Input Function estimation** | Possible using image-derived or population-based methods; more limited by small blood pool visibility per pass | More reliable estimation from large blood pools (e.g., aorta, heart chambers) always within AFOV; improved accuracy |
| **Sensitivity** | Lower sensitivity due to smaller AFOV and segmented acquisition | 4-40x higher sensitivity; enables ultra-short acquisitions |
| **Scan duration for reliable Patlak anaylsis** | Typically long (up to ~90 minutes) to achieve adequate statistics | Shorter durations feasibile (10-20 minutes sufficient in studies[ref]); compatible with standard FDG workflows |
| **Impact on PBPK modeling** | Possible but less practical; higher complexity in data stitching | Highly suitable; continuous high-quality TACs across all organs reduce uncertainties in PBPK parameter estimation |
| **Sources of error** | Higher motion sensitivity due to bed changes; more protocol complexity | Reduced motion/artifact |
| **Practicality for routine clinical workflow** | Limited by complexity, motion, and long scan times | Streamlined protocols, shorter scans, higher patient comfort; more clinically adoptable |
| **Challenges** | Long acquisition times, patient compliance, complex protocols | Data storage/processing burden; need for advanced reconstruction and analysis software |
| **Adoption Barrier** | Workflow complexity and patient burden | Infrastructure (data handling), software availability, cost considerations |